\documentclass[copyright,creativecommons]{eptcs}
\providecommand{\event}{FAVO 2009} 
\providecommand{\volume}{??} \providecommand{\anno}{2011}
\providecommand{\firstpage}{1} \providecommand{\eid}{??}
\usepackage{breakurl}        
\usepackage{graphicx}
\usepackage{amsmath}
\usepackage{xspace}
\usepackage{amssymb}
\usepackage{wasysym}
\usepackage{pifont}

\newcommand{\sAndR}{\texttt{\textbf{s\&r}}\xspace}
\newcommand{\rAndS}{\texttt{\textbf{r\&s}}\xspace}
\newcommand{\rcv}{\texttt{\textbf{rcv}}\xspace}
\newcommand{\snd}{\texttt{\textbf{snd}}\xspace}
\newcommand{\ask}{\texttt{\textbf{ask}}\xspace}
\newcommand{\rpl}{\texttt{\textbf{rpl}}\xspace}
\newcommand{\tll}{\texttt{\textbf{tll}}\xspace}
\newcommand{\prf}{\texttt{\textbf{prf}}\xspace}

\newcommand{\sndParam}{%
 \includegraphics[height=7pt]{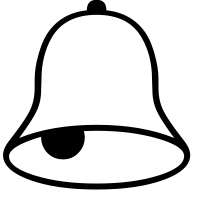}}
\newcommand{\rcvParam}{%
  \includegraphics[height=7pt]{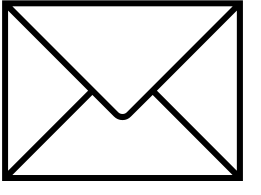}}

  \newcommand{\cmpParam}{%
  \includegraphics[height=7pt]{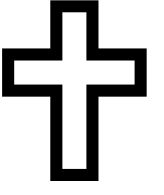}}
\newcommand{\pledge}{\includegraphics[height=7pt]{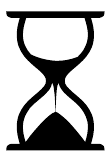}}

\newcommand{\INTERACTIONS}{\textbf{\texttt{INTERACTIONS}}}
\newcommand{\BUSINESSPROTOCOL}{\textbf{\texttt{BUSINESS PROTOCOL}}}
\newcommand{\IS}{\textbf{\texttt{is}}}

\newcommand{\BEHAVIOUR}{\textbf{\texttt{BEHAVIOUR}}}

\newcommand{\SLAVAR}{\textbf{\texttt{SLA VARIABLES}}}

\newcommand{\enables}{\textbf{enables}\xspace}
\newcommand{\until}{\textbf{until}\xspace}
\newcommand{\after}{\textbf{after}\xspace}
\newcommand{\ensures}{\textbf{ensures}\xspace}
\newcommand{\initiallyEnabled}{\textbf{initiallyEnabled}\xspace}

\newcommand{\srmlclock}{\includegraphics[height=10pt]{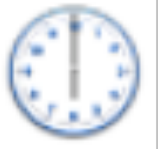}}
\newcommand{\slider}{\includegraphics[height=11pt]{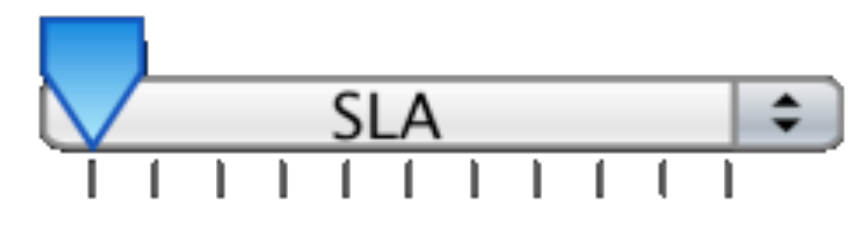}}

\title{Structure and Behaviour of Virtual Organisation Breeding Environments}

\author{
Laura Bocchi, Jos\'{e} Fiadeiro, Noor Rajper and Stephan Reiff-Marganiec
\institute{Department of Computer Science, University of Leicester\\
University Road, Leicester LE1 7RH, UK} 
\email{\{lb148,jose,nr76,srm13\}@mcs.le.ac.uk}
 }

\def\titlerunning{Structure and Behaviour of Virtual Organisation Breeding Environments}
\def\authorrunning{L. Bocchi, J.L. Fiadeiro, N.J. Rajper \& S. Reiff-Marganiec}
\def\authorcopyright{Bocchi, Fiadeiro, Rajper \& Reiff-Marganiec}
\begin{document}
\maketitle

\begin{abstract}
This paper provides an outline of a formal approach that we are developing for modelling Virtual Organisations (VOs) and their Breeding Environments (VBEs).  We propose different levels of representation for the functional structures and processes that VBEs and VOs involve, which are independent of the specificities of the infrastructures (organisational and technical) that support the functioning of VBEs.  This allows us to reason about properties of
tasks performed within VBEs and services provided through VOs without committing to the way in which they are implemented.
\end{abstract}

\section{Introduction}
This paper reports on on-going work towards a formal approach for modelling virtual organisations (VOs) and their breeding environments (VBEs) in the sense of \cite{FAVO10}. A VBE defines a base long-term cooperation agreement among
a number of participants  (individuals or institutions) and characterizes their interoperable infrastructure~\cite{FAVO09b}. As such, a VBE represents the organisational context in which the creation and operation of VOs takes place; VOs are seen as ensembles  that are formed dynamically to provide high-level functionalities, or services, by sharing a number of resources in a distributed way, using the new connectivity environments that are being made available through Global \cite{FAVO16} and Grid Computing \cite{FAVO15}.  

The purpose of developing a formal modelling approach echoes the challenge of building ``Verifiable VOs" as raised in \cite{FAVO09}.  This implies that our approach is unavoidably partial: as in any formal account of the real world (which includes business), we need to operate on abstractions that are amenable to some form of mathematical
representation and analysis.  Our approach defines different levels of representation of VBEs, VOs and their activities, which are essential for supporting several forms of analysis, from the properties of the coordination structures that are put in place through policies and workflows to the management of the resources that are shared within a VBE and used by their VOs.

There are multiple levels of abstraction at which formal methods can operate.  In this paper, we abstract from the specificities of the infrastructures (organisational and technical, including IT) that support the functioning of VBEs: we aim for an `infrastructure-agnostic' account of the functional structures and processes that VBEs and VOs involve.  We concentrate on the functional and behavioural aspects in which partners and resources are involved without committing to the way in which they are effectively implemented.  Therefore, we do not model the brokers
(human or software) that procure services that can be used to create business or the communication networks that support interconnectivity --- what we could call the VBE middleware.

We are also aware that the model that we present in this paper misses several aspects of VBEs.  For instance, we do not address at present the decision-making processes through which VOs are created within a VBE or the actual business goals that preside to the creation of a VBE (see \cite{FAVO11} for an overview of some of the formal approaches that
have been proposed to address these issues).  However, we do intend that the formal models that we propose can be used to inform such processes, for instance by supporting stochastic analysis on the usage that VOs can make on VBE resources or validation of functional properties that VOs offer through services, something that we are leaving for future work.

Our approach supports the definition of a structural and behavioural model of a fixed VBE based on three different levels of representation: (1) the definition of the \emph{persistent} functionalities of the VBE; (2) the definition of the \emph{transient} functionalities of the VOs that are offered by the VBE at a specific moment in time, what we call a \emph{business configuration} of the VBE; and (3) the ensemble of components (instances) and connectors that, at that time, deliver the services offered by the VOs present in the business configuration, what we call a \emph{state configuration}.  These levels are not `architectural layers': they do not contain entities that interact with entities in other layers.  Rather, they represent a hierarchy of representations at a fixed time: the first level is invariant, i.e. it provides a representation of those aspects of a VBE that will not change; the business configuration at the level below captures the way the VBE is logically organised at that time in terms of VOs; the state configuration represents the actual `physical' instances of the VOs that are currently operational, i.e. which specific services are currently being provided within the VBE.   `Real' entities are only represented in state configurations: the other levels deal only with types of entities.

More specifically, the three levels of representation are modelled as follows:
\begin{itemize}
\item A VBE consist of (1) a collection of \emph{resources}; (2) a consortium of (persistent) \emph{partners};
 (3)  a number of \emph{policies} constraining the way resources can be shared and
 the partners agree to do business together, including rules for the consortium to
 expand for establishing specific VOs; and (4) a number of supporting \emph{tasks} that operate processes (management or otherwise) that serve the roles enacted within the VBE.
 These constituent elements are invariant, i.e. they are present in every business configuration of the
 VBE (in the sense explained below).
\item The current business configuration of a VBE, is understood as (1) the collection of additional (non-permanent) partners, that we call \emph{associates},
and resources that are part of the VBE; (2) the tasks that support the roles of the new partners and their resources; (3) the VOs that the VBE currently supports; and
(4) the policies that apply to their instantiation and their coordination at any given time.
Tasks and VOs may rely on complementary, transient partners (which
we call `associates') that join the VBE to provide specific business
services and remain in the VBE only while those services are
required.  Associates can be fixed at VO-creation time or discovered
on the fly when needed, subject to service-level agreements, in
order to be able to accommodate the needs of specific clients.
\item The current state configuration of a VBE consists of `components', connected through `wires', that jointly operate the tasks and the services offered by the VOs that are running in the current state.  These components
include the shared resources of the VBE as well as those that are brought into the VBE by the associates. The topology of the configuration (the way components are wired together) reflects the policies established at the level of the current business configuration.  At this level, one can determine levels of resource consumption or properties of a number of other parameters, including measures of quality of service.
 \end{itemize}
Part of the importance of distinguishing between these three levels is that we can account for two different kinds of change (admitting that the VBE level is invariant, the creation of which we do not model at present):
\begin{itemize}
\item Changes in the business configuration reflect the creation or deletion of tasks or VOs. Creating a new VO may involve identifying associates or the criteria that will need to be observed for discovering such associates on the
fly, depending on the nature of the customers that procure the service (in which case each service may involve different associates).  Deleting a VO requires that the current state configuration is in a quiescent state relative to that VO, i.e. that none of the services offered by the VO is currently active. Changes at this level are triggered by business concerns (which we do not model at present).
\item Changes to the state configuration result from the launching of (instances of) tasks or of services provided by one of the VOs present in the business configuration, which dynamically adds (or removes) components or wires to (from) the current state configuration. Changes at this level are triggered by the actions performed by or
through the components and the communications exchanged through the wires that connect them.
\end{itemize}
Given the way levels are organised, these changes take place in different `timebands' in the sense of  \cite{burns}, i.e. the levels induce different granularities of time: the state-configuration changes take place within a fixed business configuration, meaning that business configurations induce a coarser time scale. 

Given the limited space available, we focus only on the VBE and business configuration levels.  In the sequel, we outline the formalisms and methodology that we are proposing for each level of representation and change, which we have adapted (and extended) from recent work on service-oriented modelling \cite{FAVO12,FAVO14}.
Essentially, we use graph-based representations to formalise and establish relationships between the two levels --- logic/process-based formalisms for the specification of activities and services.

\section{A Model of Virtual Organisation Breeding Environments}
As already mentioned, we see a VO as a dynamic ensemble of entities that operate over a communication and collaboration network through which they can share resources to offer services.  Some of those entities and
resources are provided by the VBE in which the VO was created; others are external to the VBE and co-opted or procured to satisfy the business goal of that particular VO.  We define a (formal model of a) VBE to consist of:
\begin{itemize}
\item A collection of persistent partners, where a partner consists of:
\begin{itemize}
\item Its name (individual or organisation, virtual or not);
\item A collection of attributes through which policies can be defined on the involvement of the partner in business configurations.
\end{itemize}
\item A collection of persistent resources where a resource consists of:
\begin{itemize}
\item Its identifier;
\item  A collection of attributes through which the usage of the identified resource can be
monitored.
\end{itemize}
\item A collection of policies expressed over partners and resources that apply to all business configurations of
the VBE.
\item A collection of tasks that support the roles enacted within the VBE.  A task is defined by a \emph{task-module} consisting of:
\begin{itemize}
\item Component specifications that are used in state configurations as interfaces to the partners (in which they are called \emph{serves-interfaces}) or resources (in which they are called \emph{uses-interfaces}) involved in the
task;
\item Specifications of components and wires that jointly orchestrate the task.
\item Mappings from the \emph{serves-interfaces} (resp. \emph{uses-interfaces}) to the partners (resp. resources) of the
VBE complete the task definition.
\end{itemize}
\end{itemize}

Notice that we separate the \emph{task-module} from the way it is used in the VBE. Effectively, task-modules are design primitives that define patterns that can be reused in the definition of VBEs.

As an example, we use a very simple scenario: a group of hotels, a car rental company and a guided-tour company
decide to create a VBE --- \emph{visitUs} --- to promote tourism in the local town.
\begin{itemize}
\item The partners of \emph{visitUs} are the hotels (to make the example shorter, we model the case of two hotels, \emph{grandHO} and \emph{centralHO}), the car rental agency \emph{carHI} and the guided-tour company \emph{tourAG};
\item The resources include two systems: \emph{registrationSY} supporting management activities of \emph{visitUs} and a shared reservation system \emph{reservationSY} supporting the business purpose (hotel bookings and so on).
\item The policies establish criteria for the admission of transient partners (e.g. they need to operate
within a given vicinity) and the use of the shared resources (e.g. the cost of maintaining the reservation system), for which their interfaces need to include parameters that capture these properties. The policies are expressed as first-order expression in the language of the parameters and the corresponding data types.
\item The tasks include the process \emph{managerRO} that supports the administrator role performed by \emph{grandHO}, which connects to the registration system, and the process \emph{memberRO} that allows each partner to use the reservation system.  (Other roles might have been considered as discussed in \cite{FAVO11}.)
\end{itemize}

\begin{figure}[h]
  \begin{center}
 \includegraphics[height=5.00cm]{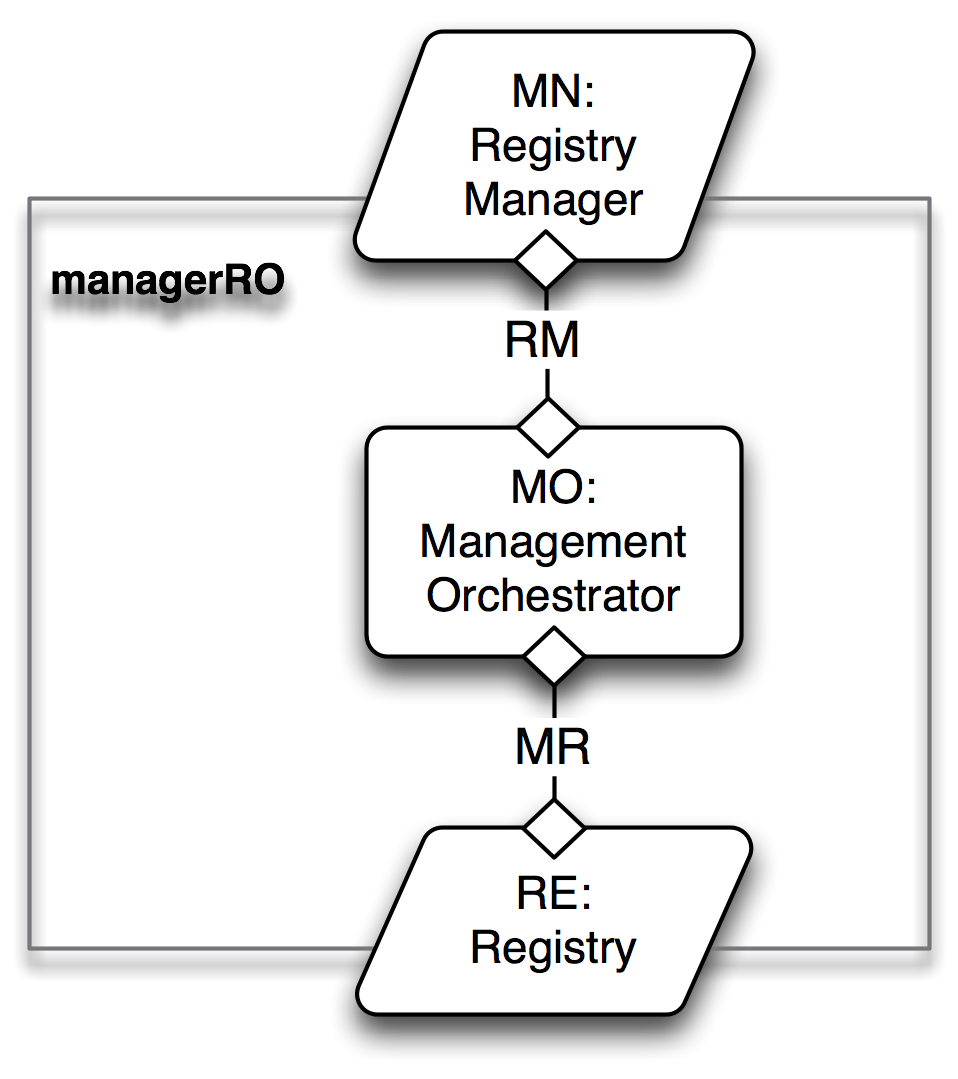}
\end{center}
 \caption{The task module \emph{managerRO}}
 \label{fig1}
 \end{figure}

We use a graphical notation to depict task modules as illustrated in Figure \ref{fig1} for the task \emph{managerRO} that supports the administrator of \emph{visitUs}.  The specification of the component MO that orchestrates the task is Management Orchestrator and the wires are \emph{RM} and \emph{MR}. The specification of the serves-interface \emph{MN} used by \emph{grandHO} - the partner who performs the managerial role - is Registry Manager, and the specification of the uses-interface \emph{RE} (which connects to \emph{registrationSY}, the resource that supports the task) is Registry.

\begin{figure}[h]
  \begin{center}
 \includegraphics[height=8.00cm]{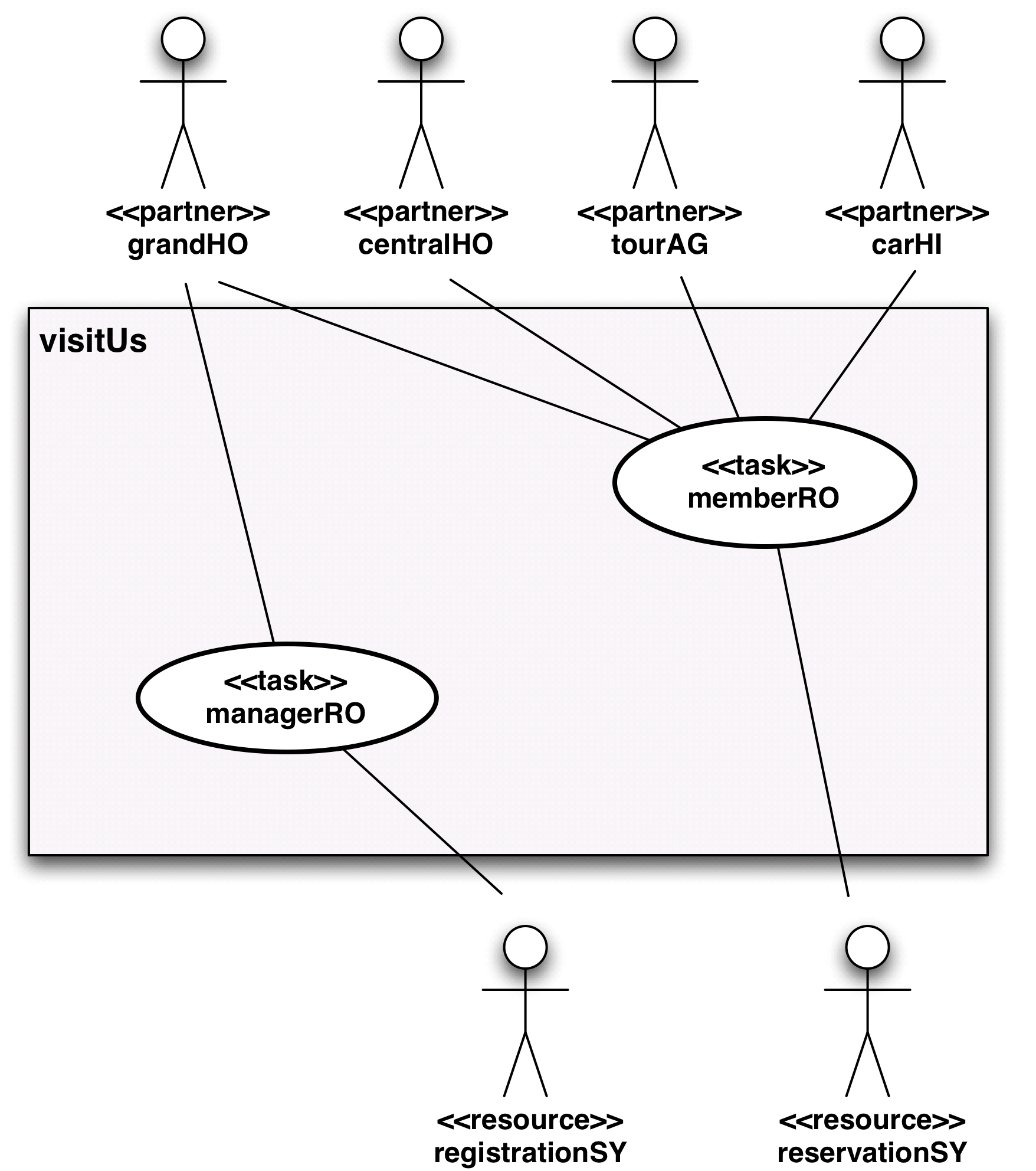}
\end{center}
 \caption{VBE diagram for \emph{visitUS}}
 \label{fig2}
 \end{figure}

We also use a graphical notation to depict VBEs, which is inspired by use-case diagrams (though our usage of the notation is not necessarily faithful to its original purpose).  As illustrated in Figure \ref{fig2}, we use stereotypes to identify the actors that correspond to partners and resources of the VBE.  Each task is represented by a use-case.

\section{Business Configurations}
The current business configuration of a VBE defines the types of tasks and the VOs that the VBE supports to meet its (current) business goals.  The actual process through which a VBE decides on how to configure itself is not part of our present model as it depends on a number of business or organisational concerns that the formal methods that we are illustrating do not address.  However, the kinds of qualitative and quantitative analysis that our approach provides should be able to inform that process and corresponding decisions, something that we plan to address in the future. 

We define a VBE business configuration to consist of an extension of the VBE with:
\begin{itemize}
\item A collection of \emph{associates} (transient partners), where an associate consists of:
\begin{itemize}
\item Its name (individual or organisation, virtual or not);
\item A collection of attributes through which policies can be defined on the involvement  of the associate in the VBE.
\end{itemize}
\item A collection of transient resources, each of which consists of:
\begin{itemize}
\item Its identifier;
\item A collection of attributes through which the usage of the identified resource can be monitored.
\end{itemize}
\item A collection of policies expressed over associates and their resources that apply to their involvement in the VBE (e.g. the conditions that determine the cessation of their involvement).
\item A collection of tasks that support the roles enacted by the associates within that business configuration of the VBE.
\item A collection of VOs that define the services that the VBE provides in that business configuration. A VO is defined by a \emph{VO module} consisting of:
 \begin{itemize}
 \item Component specifications that are used in state configurations as serves-interfaces to the partners or 
 associates, or uses-interfaces to resources involved in the VO; typically, one such interface serves the
 coordinator of the VO.
 \item Component specifications that are used as requires-interfaces for external entities or as the provides-interface for the customers of the VO.  The specification of requires-interfaces identifies the behavioural properties that are expected of external parties to be eligible to be chosen as service providers for the VO.  The specification of the provides-interface identifies the properties that customers can expect of the service offered by the module.
  \item Specifications of the components and wires that model the (possibly distributed) process that orchestrates the services provided by the VO.
  \item  An internal configuration policy, which identifies the triggers of the external service discovery process as well as the initialization and termination conditions of the components and wires.
\item An external configuration policy, which consists of the variables and policies that determine the quality profile to which the discovered services need to adhere.
\item Mappings from the serves-interfaces (resp. uses-interfaces) to the partners or associates (resp. resources) of the VBE complete the VO definition.

 \end{itemize}
\end{itemize}

\begin{figure}[h]
 \begin{center}
  \includegraphics[height=7.00cm]{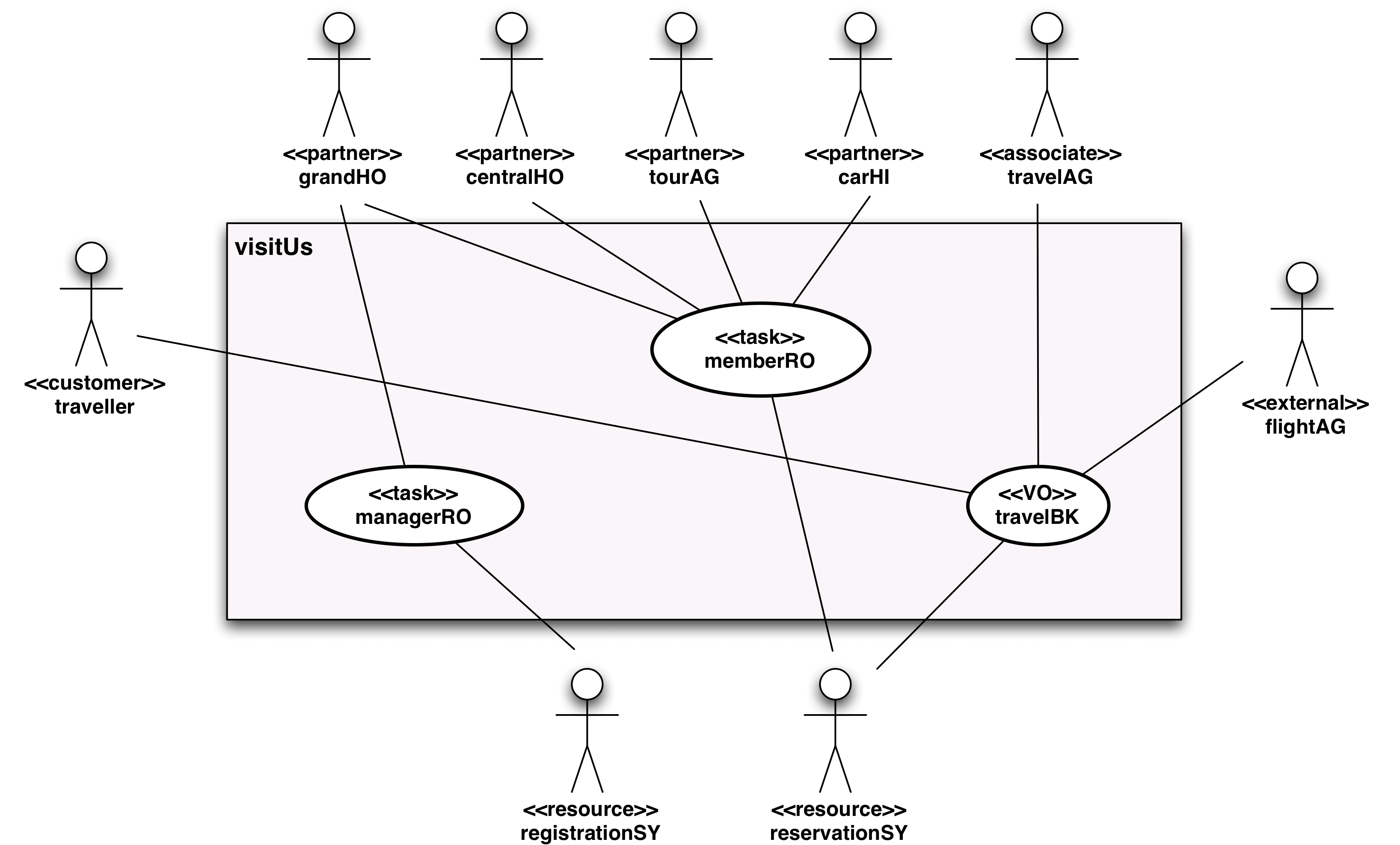}
 \end{center}
  \caption{VBE business configuration for \emph{visitUS}}
  \label{fig3}
  \end{figure}

\begin{itemize}
\item A collection of \emph{external entities}, each of which represents a partner that may need to be co-opted to provide a service for one of the VOs that the VBE offers in that business configuration.
\item A collection of \emph{customers}, one for each of VO, each of which defines the interface (interactions and functional properties) that the customer of the corresponding VO can expect.
\end{itemize}
As for tasks, VO-modules are design primitives that define patterns that can be reused in the definition of multiple VBE business configurations. As an example, we consider a business configuration of \emph{visitUs} in which a travel booking service is offered through a VO named \emph{travelBK}. An associate named \emph{travelAG} is admitted as a member of the VBE for managing that VO. Services offered through \emph{travelBK} may require an external flight agent to be discovered according to the criteria specified in \emph{flightAG}. A specific agent is not chosen as an
associate in order to maximise customer satisfaction --- each customer of the VO may express service-level policies (e.g. preference for a particular airline, or minimum cost, or proximity) that will be optimised when selecting the corresponding external partner.
\newline We extend the diagrams used for VBEs to account for business configurations as illustrated in Figure \ref{fig3}.

\begin{figure}[h]
  \begin{center}
 \includegraphics[height=5.00cm]{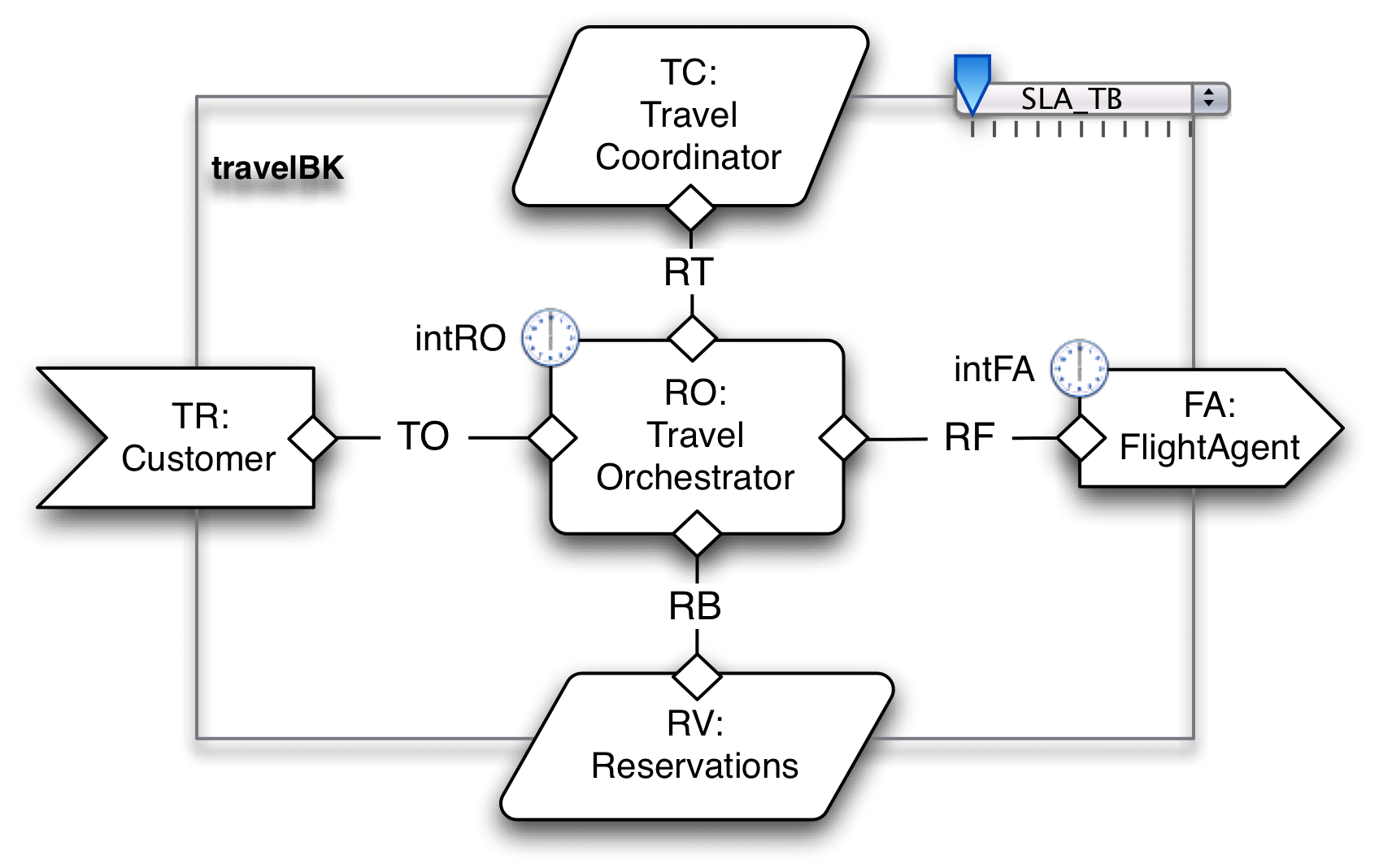}
 \end{center}
  \caption{The VO module \emph{travelBK}}
  \label{fig4}
  \end{figure}

We use a graphical notation similar to task-modules to depict VO-modules as illustrated in Figure \ref{fig4} for
\emph{travelBK}. We use the symbol \srmlclock\ to indicate the internal configuration policy as it applies to components and requires interfaces, and \slider\ for the external configuration policy.  The module consists of a provides-interface \emph{TR} for interactions with customers of the VO, a serves-interface for interactions with the coordinator of the VO, a uses-interface for interactions with the reservations system, and a requires-interface for the discovery of a flight agent.\newline

A more formal definition follows where each node uniquely represents a specific interface of an entity (e.g., institution, participant, etc.) in a business configuration rather than the entity itself.
A task- or VO-module \emph{M} defines:
\begin{itemize}
\item A graph \emph{graph(M)}.
\item A distinguished subset of nodes \emph{uses(M)}$\subseteq$ \emph{nodes(M)}.
\item A distinguished subset of nodes \emph{serves(M)} $\subseteq$ \emph{nodes(M)} distinct from \emph{uses(M)}.
\item In the case of a VO-module, a subset of nodes \emph{requires(M)} $\subseteq$ \emph{nodes(M)} distinct
from \emph{uses(M)} and \emph{serves(M)}.
\item In the case of a VO-module, a node \emph{provides(M)}$\in$ \emph{nodes(M)} distinct from
\emph{requires(M)}, \emph{serves(M)} and \emph{uses(M)}.
\item A labelling function \emph{$label_M$} such that:
\begin{itemize}
\item \emph{$label_M$(n)} is a \emph{component specification}.
\item \emph{$label_M$ ($e:n$ $\leftrightarrow$ m)} is a \emph{connector}.
\end{itemize}
\end{itemize}
Component specifications and connectors are discussed in Section \ref{sec4}. In the case of a VO-module, we denote by
\emph{body(M)} the (full) sub-graph of \emph{graph(M)} that forgets the node \emph{provides(M)}, the nodes in \emph{requires(M)} and the edges that connect them to the rest of the graph.  That is, \emph{body(M)} consists of all the elements that are internal to the VO. 

\begin{figure}[h]
  \begin{center}
  \includegraphics[height=8.00cm]{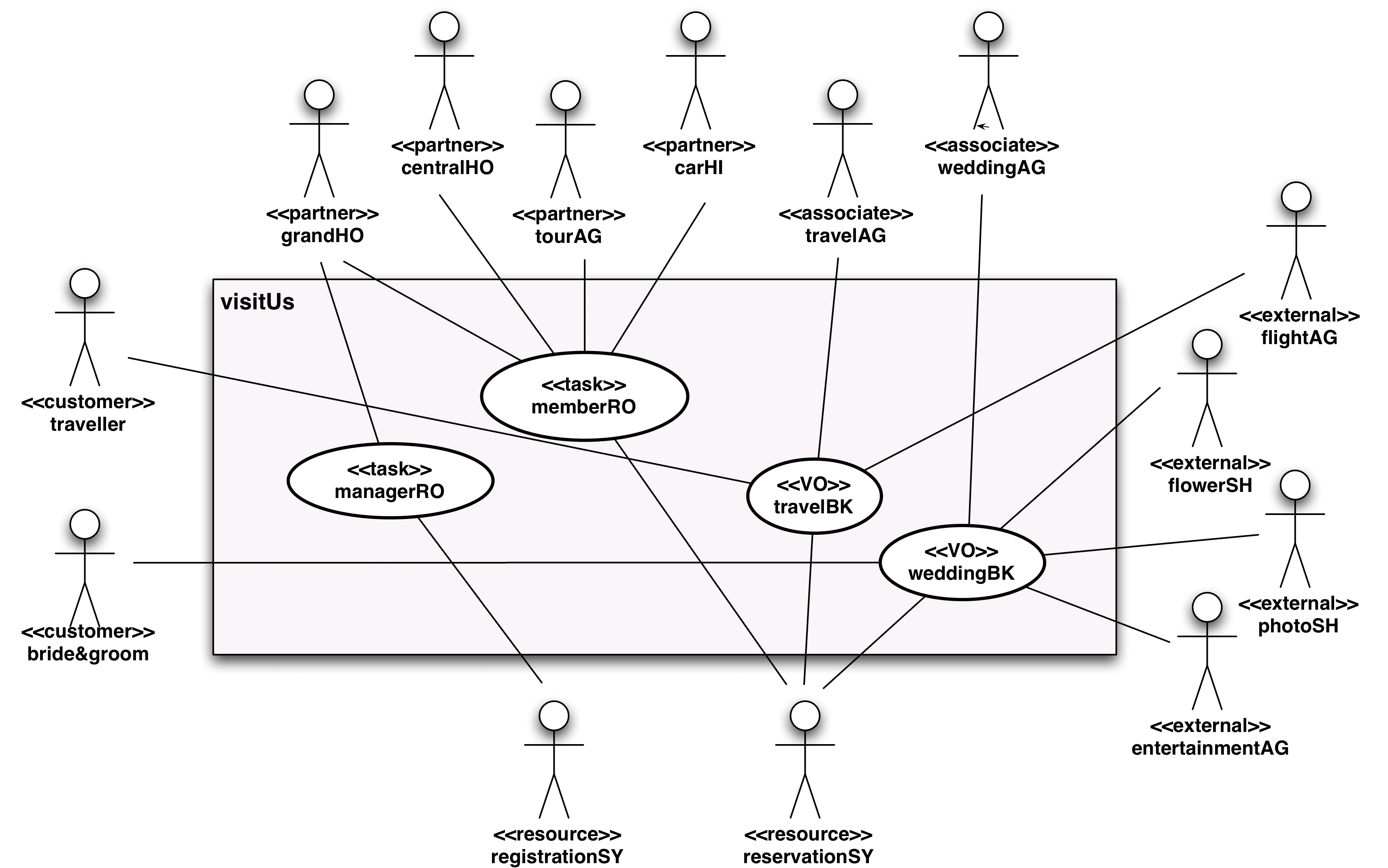}
 \end{center}
  \caption{Another VBE business configuration for  \emph{VisitUS}}
  \label{fig5}
  \end{figure}

A business configuration of a VBE also defines a labelled graph obtained by expanding its tasks and VOs with the bodies of the labelled graphs that correspond to their modules.  Having such a formal representation for VBE configurations allows us to use graph transformations to formalise rules and policies to evolve the configurations, for instance the creation of new VOs.  In Figure \ref{fig5}, we depict a business configuration that extends the one
in Figure \ref{fig3} with a new VO that offers arrangements for weddings as a service.

\section{Component Specifications and Connectors}
\label{sec4}
In order to account for the behaviour that, in state configurations (referred to as level 3 in Section 1), emerges from the interconnections established inside the ensembles that perform tasks or deliver services through VOs, we need a uniform representation of the entities and resources involved, which in our approach we do in terms of component and wire specifications.  
A component specification is a pair $\langle$\emph{signature}, \emph{behaviour}$\rangle$ where:
\begin{itemize}
\item \emph{Signature} declares the interactions in which the component may be involved.
\item \emph{Behaviour} is a formal model of the behaviour of the entity that the component represents
expressed in terms of the interactions identified in the signature and a number of parameters that reflect
resource consumption or quality-of-service attributes.
\end{itemize}
Given the space available, we are not able to define in detail the formalisms that we use in component specifications (these are similar to those that we have proposed for the service modelling language SRML \cite{FAVO12}).  We discuss below the provides-interface of \emph{travelBK}, which is of type \emph{Customer}.  This specification is what we call a \emph{business protocol}: it uses patterns of typical business conversations, which are abbreviations of sentences of a temporal logic that we have adopted for service-oriented modelling \cite{FAVO03}.  The remaining specifications can be found in the Appendix.
 
In the formalism that we adopt, interactions can be either synchronous or asynchronous, one-way or two-way (i.e.
conversational): 
\begin{itemize}
\item
\sAndR/\rAndS\ --- conversational asynchronous interaction where the initiating party expects
a reply from its co-party
\item
\rcv/\snd\  --- one-way asynchronous
receive/send
\item
\ask/\rpl\  --- synchronization with the co-party to obtain/transmit data
\item
\tll/\prf\  --- blocking requests to the co-party to perform an operation
\end{itemize}
In our example, the interface that \emph{travelBK} offers to its customers specifies that the VO can engage in the interaction \emph{bookTrip} (initiated by the customer) and send \emph{payNotify} and \emph{refund} to the customer.

\begin{center}
\begin{footnotesize}
\[  \begin{array}{l}
\qquad \BUSINESSPROTOCOL\ \ \mathit{Customer} \ \ \IS \\
 \qquad \quad \INTERACTIONS\\
\qquad \quad \qquad \rAndS\ \mathit{bookTrip}\\
\qquad \quad \qquad\quad \sndParam ~~\quad \mathit{from,to}\,:\,\mathit{airport},  ~~\mathit{out,in}\,:\,\mathit{date}\\
  \qquad \quad \qquad \quad  \rcvParam \quad \mathit{fconf}\,:\,\mathit{fcode}, ~~\mathit{hconf}\,:\mathit{hcode},~~\mathit{amount}\,:\mathit{moneyvalue}\\
 \qquad \quad \qquad \snd ~~\mathit{payNotify}\\
\qquad \quad \qquad\quad \rcvParam \quad \mathit{status}\,:\,\mathit{bool}\\
 \qquad \quad \qquad \snd ~~\mathit{refund}\\
\qquad \quad \qquad\quad \rcvParam \quad \mathit{amount}\,:\,\mathit{moneyvalue}\\
 \qquad \quad  \SLAVAR\\
 \qquad \quad \qquad KD:[0..100], PERC:[0..100]\\
 \qquad \quad \BEHAVIOUR\\
 \qquad \quad \qquad \initiallyEnabled~bookTrip~\sndParam ?\\
 \qquad \quad \qquad (bookTrip~\pledge \wedge bookTrip~\checkmark?)~\ensures~payNotify~\sndParam!\\
 \qquad \quad \qquad (payNotify~\sndParam! \wedge payNotify.status)~\enables~bookTrip~\cmpParam?~\until~today+KD<bookTrip.out\\
 \qquad \quad \qquad (bookTrip~\cmpParam? \wedge today+KD<bookTrip.out)~\ensures~refund~\sndParam!\\
 \qquad \quad \qquad refund.amount > bookTrip.amount * PERC/100~\after~refund~\sndParam!\\
 \end{array}\]
\end{footnotesize}
\end{center}

Interactions
of type r\&s and s\&r are conversational (in the sense of
\cite{FAVO06}), i.e. they involve a number of events exchanged
between the two parties:

\small
\begin{scriptsize}
\begin{table}[h]
\centering
\begin{tabular}[h]{| p{1.0in} | p{3.0in} |}
 \hline
 \emph{interaction\sndParam} & The event of \emph{initiating interaction.} \\

 \hline
 \emph{interaction\rcvParam} & The reply-event of \emph{interaction}. \\

 \hline
 \emph{interaction\checkmark} &  The commit-event of \emph{interaction}. \\

 \hline
 $\emph{interaction\ding{56}}$  & The cancel-event of \emph{interaction}. \\

 \hline
 \emph{interaction\cmpParam} & The revoke-event of \emph{interaction}.\\

 \hline
\end{tabular}
\end{table}
\end{scriptsize}
\normalsize

The meaning of these events should be self-explanatory: the \emph{reply-event} is sent by the co-party, offering a deal or declining to offer one; in the first case, the party that initiated the conversation may either commit to the deal or cancel the interaction; after committing, the party can still revoke the deal, triggering a compensation mechanism.  Events can have several parameters (for instance, the initiation event \emph{bookTrip}\sndParam\xspace carries data about airports and dates), and the corresponding reply event \emph{bookTrip}\rcvParam\xspace carries reservation codes for the flight and the hotel as well as the total cost).
 
These events are used as atomic formulae in the language that we use to specify the properties that a customer can expect from the service.  For instance, the first property specifies that the \emph{VO} is ready to receive the initiation event of \emph{BookTrip}. The second property says that a commit event received during the validity period of the booking entitles the customer to receive a pay confirmation.

The declaration of the interactions in a signature is local to the component, i.e. all interaction names are local.  This implies that there are no implicit relationships between components that result from the accidental of the same name: all interconnections are externalised instead in what we call `wires'.  A wire defines a connector through which two components can be interconnected so that they can interact.  More specifically, a connector \cite{FAVO02,FAVO19} is a triple $\langle$\emph{roleA,Glue,roleB}$\rangle$ where \emph{roleA} and \emph{roleB} are signatures and Glue defines the protocol that coordinates the interactions identified in \emph{roleA} and \emph{roleB} --- this may include routing events, superposing protocols for secure communication, or transforming sent data to the format expected by the receiver, inter alia.  A wire interconnects two components through the connector by mapping \emph{roleA} to one component and \emph{roleB} to the other.

Service-level agreements are negotiated through policies using the c-semiring approach to constraint satisfaction
and optimisation ~\cite{FAVO07}.  An example of a policy is:\\
\indent \indent \{TC.KD,TR.PERC\}
\[
    \emph{def$_1$(d,p)} =
    \begin{cases}
      1 & \ \emph{if d $\in$ [0..100] and 1 $\leq$ d and p $\leq$ 90 and p$\leq$50+5*d }

      \\
      0, & \emph{otherwise}
    \end{cases}
  \]

The policy expresses that percentage p of the cost that is refundable (transmitted to the customer through the SLA variable \emph{TR.PERC}) is bounded by the least of 90\% and a linear function of the period d during which the deal can be revoked, which is established by the VO coordinator through the variable \emph{TC.KD}.

\section{Concluding Remarks and Further Work}
In this paper, we have outlined a formal approach that we are defining for modelling structural and behavioural aspects of VBEs and VOs.  Several levels of representation are proposed for VBEs that distinguish between (1)  the persistent aspects of VBEs in terms of members, resources and tasks that involve them, (2) the possible business configurations of VBEs characterised in terms of the VOs that it creates to provide services and the additional (associate) members that are involved in the VOs, and (3) the state configurations of VBEs, which result from the
services (instants) offered by the VOs at a given state.

From a formal point of view, these levels of representation are graphs whose nodes are component
specifications and the edges (wires) are connectors.  Component specifications provide either interfaces for partners and resources to be involved in tasks and services offered through VOs, or orchestrations of those services, or requirements for external services, or properties offered to customers of VOs.  Choosing graphs as formal models allow us to use techniques that have been proposed for formalising architectural aspects of system structure
and evolution (e.g. \cite{FAVO14,FAVO18}) in order to account fort he evolution of VBE business configurations (as VOs are added, deleted or modified) and also their configuration states (as new services are created and bound to customers). 

As formalisms for specification, we are using those put forward for service-oriented modelling in the SENSORIA project
\cite{FAVO02,FAVO03,FAVO12}. Together with the graph-based representation of business configurations, these formalisms can be used for inferring emergent properties of VOs. Model-checking techniques have been used for verifying properties offered by services \cite{FAVO04}, which we plan to extend to VOs.  The proposed formal model also supports forms of quantitative analysis using the stochastic analyser PEPA \cite{FAVO08,FAVO17}, which we intend to extend to VOs. Negotiation of service-level agreements is supported by techniques for constraint optimisation \cite{FAVO07},
which again we plan to use for the discovery of services from external partners that VOs may require.

\newpage
\bibliographystyle{eptcs}

\newpage
\includegraphics{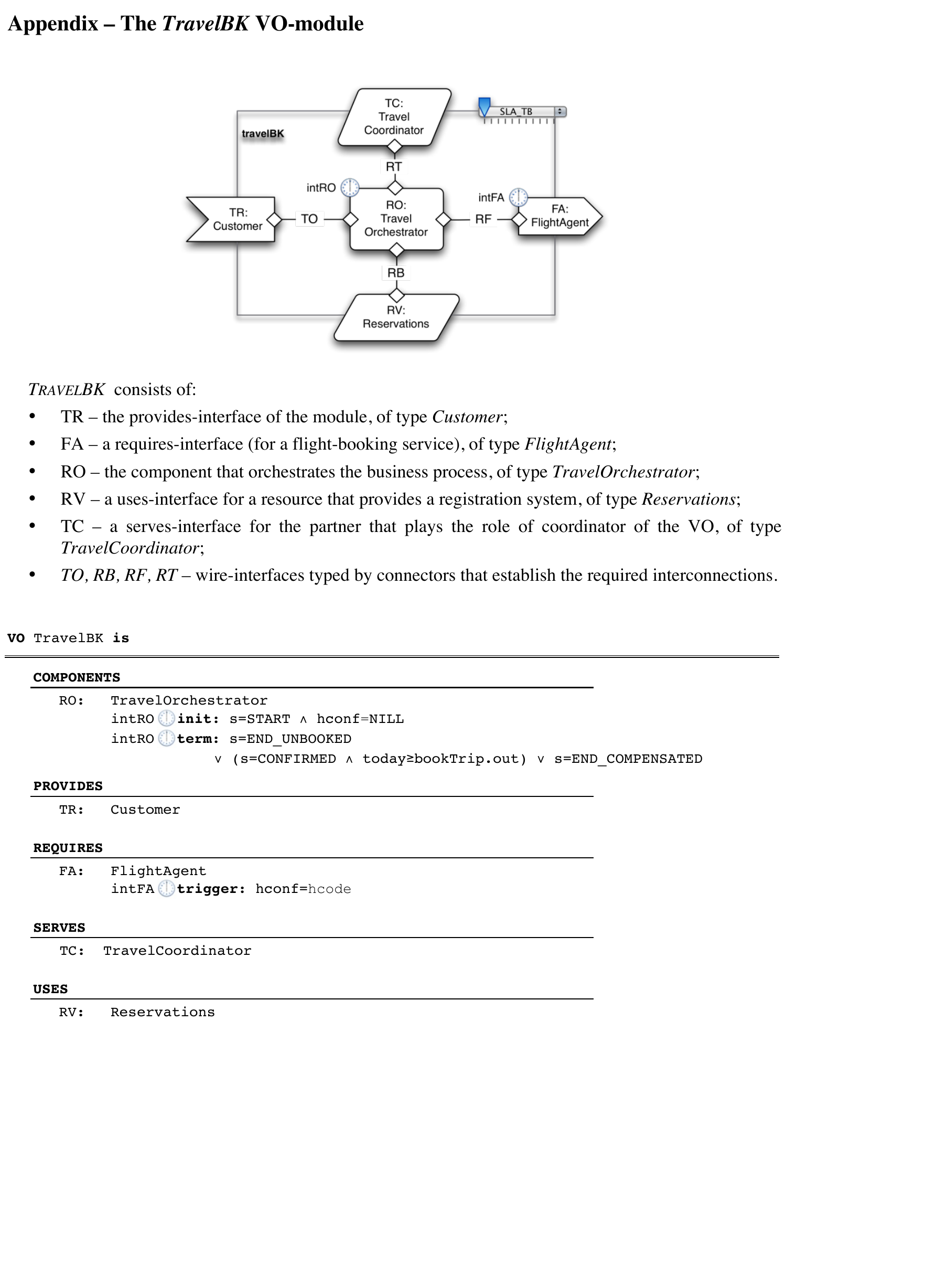}
\newpage
\includegraphics{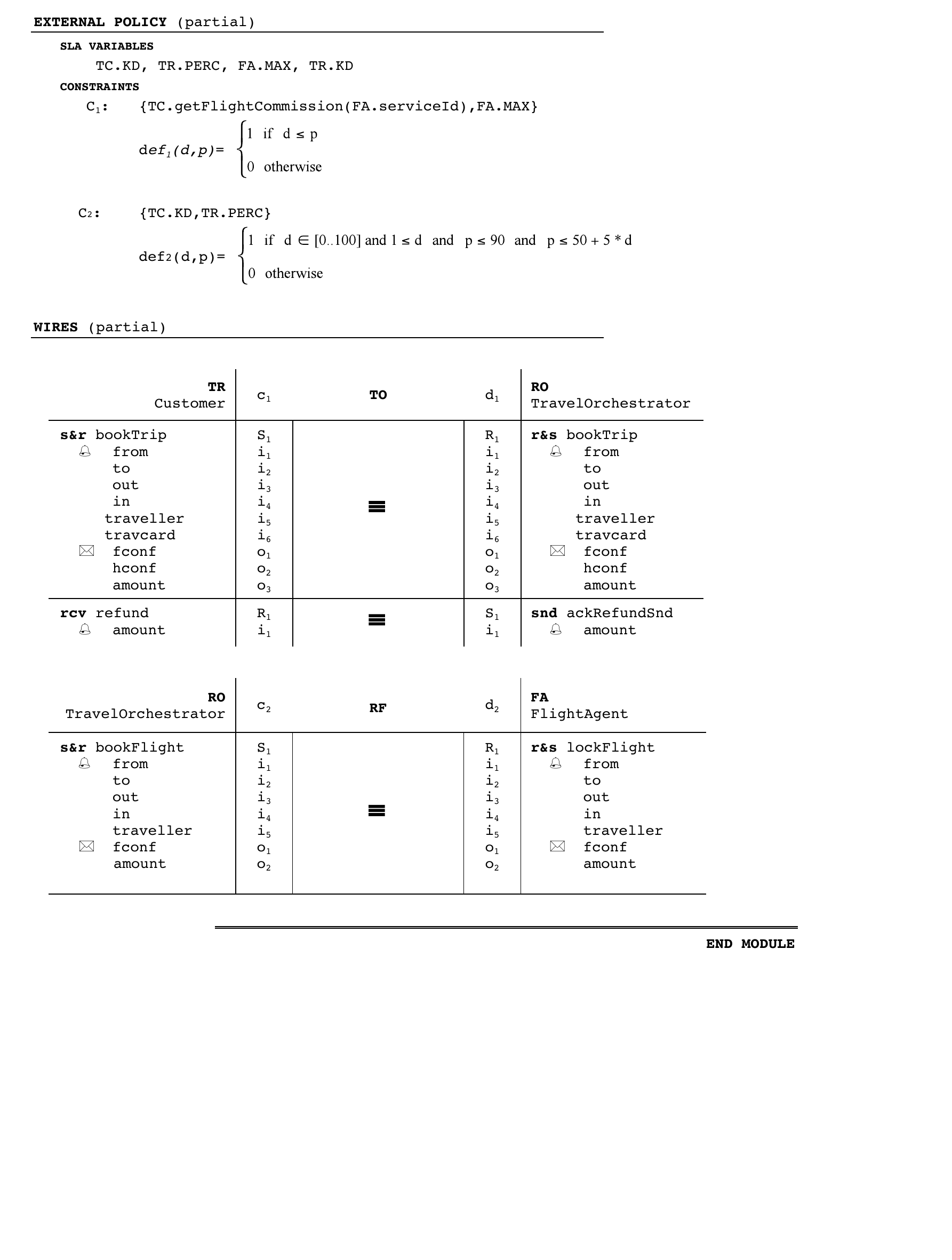}
\newpage
\includegraphics{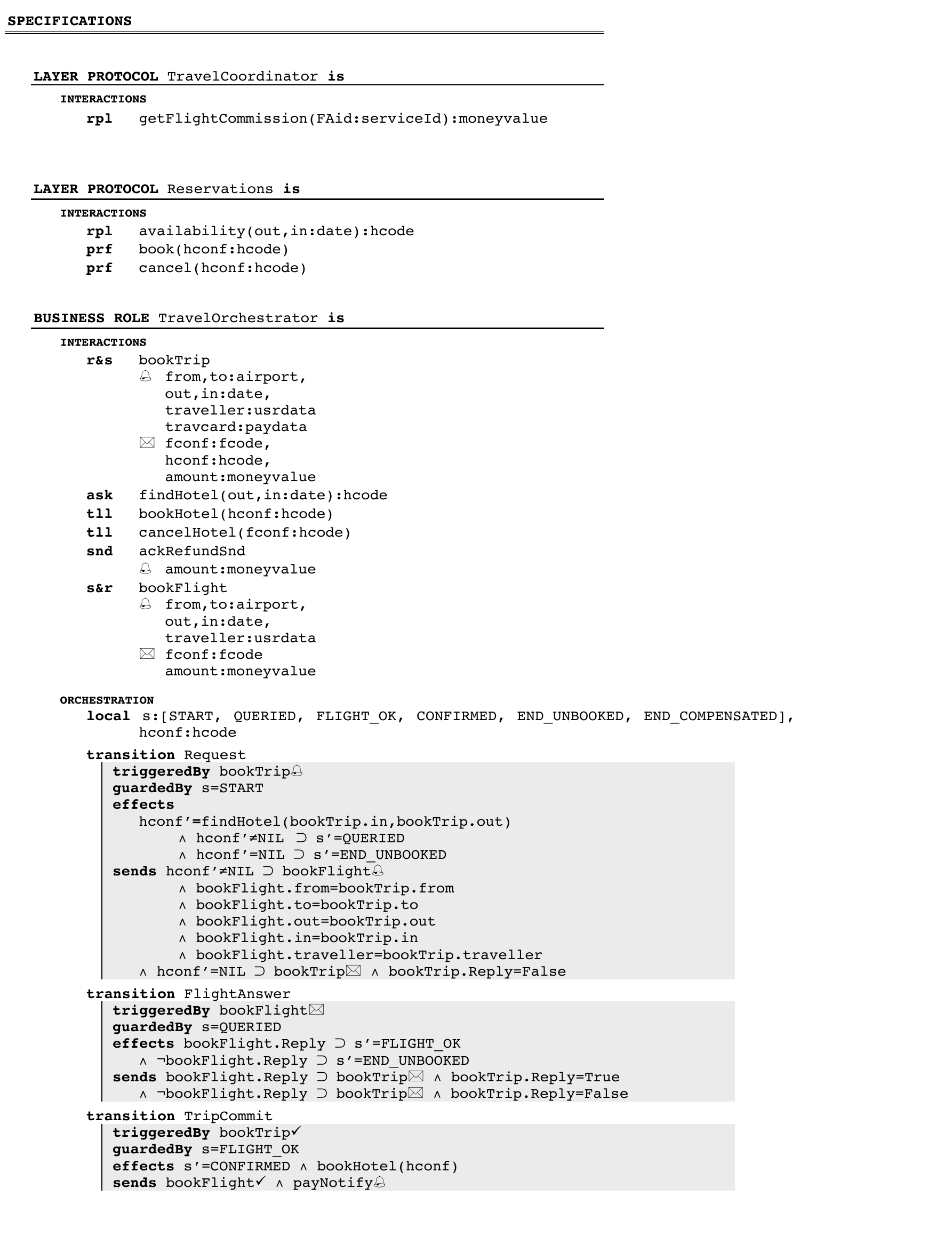}
\newpage
\includegraphics{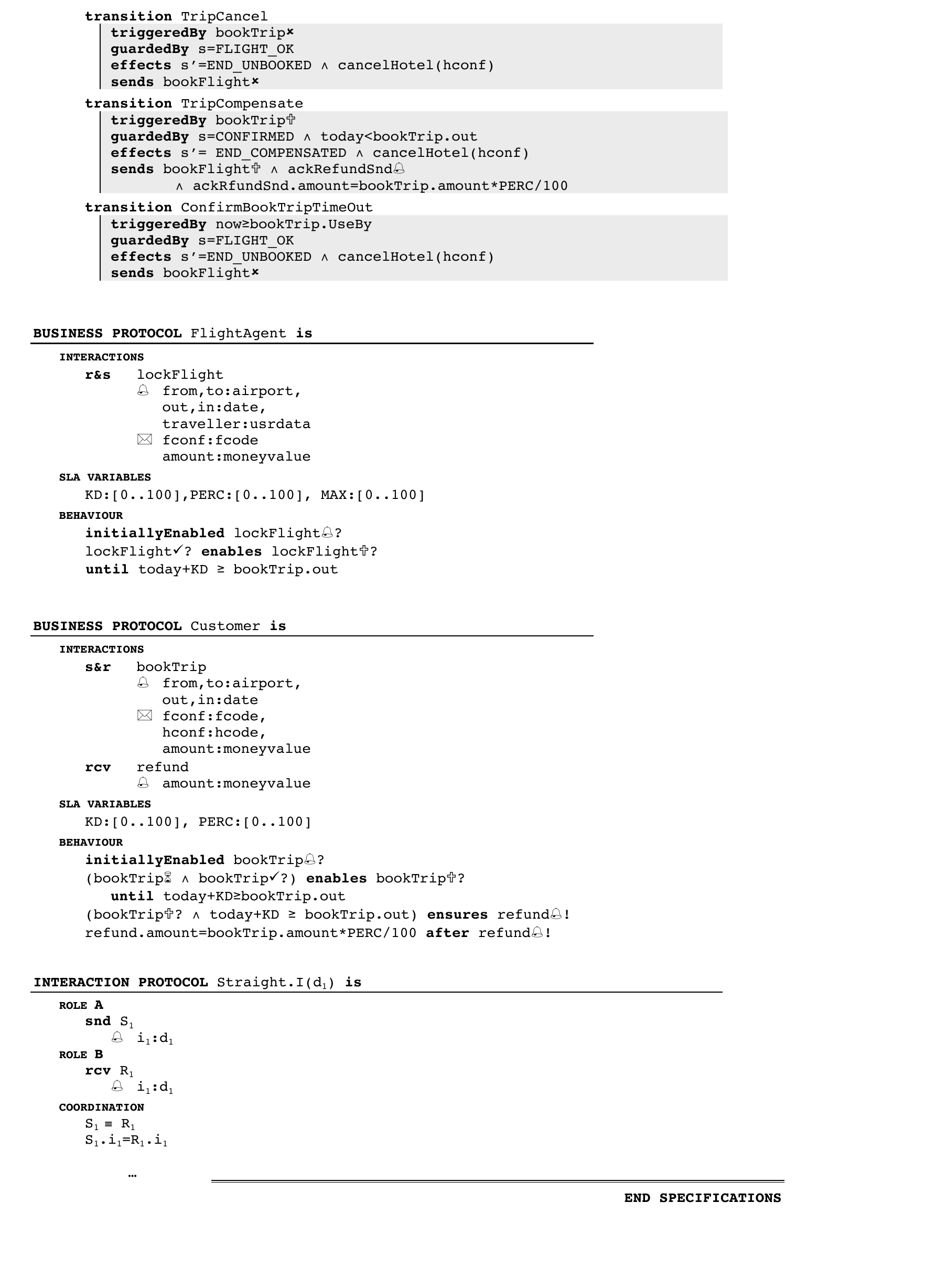}

\end{document}